\documentclass[num-refs]{wiley-article}

\papertype{Original Article}
\paperfield{Soft Matter / Biophysics}

\title{Programming Nonlinear Interfacial Mechanics of Synthetic Cells: Lipid Geometry and DNA Nanostructures}

\author[1]{Kazutoshi Masuda}
\author[1,2,3]{Miho Yanagisawa}

\affil[1]{Komaba Institute for Science,
Graduate School of Arts and Sciences,
The University of Tokyo, Komaba 3-8-1, Meguro, Tokyo 153-8902, Japan}
\affil[2]{Department of Physics, Graduate School of Science, The University of Tokyo, Hongo 7-3-1, Bunkyo, Tokyo 113-0033, Japan}
\affil[3]{Center for Complex Systems Biology, Universal Biology Institute, The University of Tokyo, Komaba 3-8-1, Meguro, Tokyo 153-8902, Japan}

\corraddress{Miho Yanagisawa, Kazutoshi Masuda}
\corremail{\\
(M. Y.) myanagisawa@g.ecc.u-tokyo.ac.jp\\
(K. M.) kazumassuu@g.ecc.u-tokyo.ac.jp}

\runningauthor{Masuda et al.}

\begin{document}

\maketitle

\begin{abstract}
Soft interfaces formed by lipid membranes are fundamental to living cells, synthetic cells, and membrane-based soft materials. However, a quantitative framework linking molecular organization with nonlinear interfacial mechanics remains elusive. Here, we establish an analytical framework that captures the nonlinear elastic response of lipid-membrane-coated synthetic cells under micropipette aspiration. Incorporating both area stretching and curvature bending enables the model to quantitatively reproduce the complete pressure–displacement response within the small-deformation regime. This approach reduces interfacial mechanics to two parameters: the in-plane area-stretching modulus and an out-of-plane bending-related term.
Using this unified framework, we experimentally demonstrate that nonlinear interfacial mechanics can be programmed by altering the molecular geometry and effective dimensionality of adsorbed elements. The lipid molecular shape and curvature-dependent packing regulate in-plane stiffness, while DNA nanostructures, the other adsorbed element, introduce an orthogonal control axis via dimensionality: isolated motifs primarily enhance area stretching, whereas three-dimensional network architectures markedly reinforce bending resistance. Together, these results establish a general molecular design principle for programming interfacial mechanics and provide a quantitative foundation for engineering mechanically tunable synthetic cells and soft interfaces.
\end{abstract}



\section{Introduction}
Soft interfaces formed by lipid membranes are fundamental building blocks of living cells, synthetic cells, and\\ membrane-based soft materials \cite{Discher2002, Sackmann1996, Seifert1997}. In biological systems, lipid membranes provide structural integrity while enabling dynamic mechanical regulation and biochemical activity within cells \cite{Lingwood2010}. Beyond biology, lipid-coated droplets and liposomes have emerged as versatile platforms for synthetic cells, soft microcapsules, and biomedical applications, including drug delivery capsules \cite{Dimova2019}. Across these systems, functionality is closely tied to mechanical response, positioning lipid-based interfaces as a prototypical class of soft matter characterized by nonlinear interfacial mechanics.

Despite their broad relevance, a quantitative understanding of how molecular organization controls nonlinear interfacial mechanics remains limited. Interfacial mechanical properties do not arise solely from the chemical identity of lipids, but instead emerge from collective molecular features such as geometry \cite{Gudyka2024, Jin2021}, packing density \cite{Kim2003, Tsang2017, Mangiarotti2025}, and the presence of additional components localized at or beneath the interface \cite{Santinho2021, Chrysanthou2024}. DNA nanostructures are particularly attractive in this context, as their geometry and connectivity can be programmed with molecular precision \cite{Bujold2018, Shen2020}. This capability opens new opportunities for material design \cite{Kaletta2025}, while simultaneously raising a fundamental question that spans soft-matter physics, biophysics, and materials science: how do the geometry and effective dimensionality of adsorbed molecular elements determine the emergent mechanics of soft interfaces?

Addressing this question requires both reliable experimental access and a physically based framework of interpreting deformation. Micropipette aspiration is a powerful method for probing the mechanics of soft, cell-like systems by directly relating applied pressure to interfacial deformation ~\cite{Rand1964, Hochmuth2000, Lee2014}. 
Although theoretical descriptions exist for large deformations, whether static or time-dependent~\cite{Waugh1979,Guevorkian2010}, analytical models that capture small yet nonlinear deformations remain limited. In practice, experimental aspiration data are often analyzed using conventional models, such as the half-space model~\cite{Theret1988} or the classic Young--Laplace law~\cite{Evans1990}, which estimate elastic moduli or interfacial tension by fitting only the linear regime of the deformation process. Therefore, the pronounced nonlinear response observed at minimal deformations is typically neglected.
A key limitation of these conventional models is that they do not explicitly incorporate not thermal but curvature-induced bending elasticity, despite its central role in theoretical and computational descriptions of soft interfaces, including lipid membranes~\cite{Evans1997, Henriksen2004}. Such approaches fail to capture the interfacial mechanics of lipid membranes with intrinsic molecular curvature or those of membranes functionalized with nanostructures possessing finite rigidity. Consequently, existing analytical frameworks remain insufficient for quantitatively describing the early nonlinear mechanical response of lipid-based soft interfaces.

Here, we introduce a physics-based analytical framework that quantitatively captures the nonlinear deformation of lipid-membrane-coated synthetic cells under aspiration. By integrating area stretching and curvature elasticity, the model reproduces the full pressure--displacement response before large deformation with viscous flow. It reduces interfacial mechanics to two physically meaningful parameters: an in-plane area-stretching modulus and an out-of-plane bending-related term. Using this unified framework, we demonstrate that nonlinear interfacial mechanics can be rationally programmed through a common physical principle: the molecular geometry, effective packing density, and dimensionality of adsorbed elements. Lipid geometry and packing density selectively regulate in-plane stiffness. In contrast, DNA nanostructures introduce an orthogonal control axis via their dimensionality, with isolated motifs enhancing area stretching and three-dimensional networks reinforcing bending resistance. Together, these results establish a general strategy for engineering mechanically programmable soft interfaces and synthetic cells.

\section{Analytical model for nonlinear elastic response}

\subsection{Energy change of elastic shells due to pipette aspiration deformation}

\begin{figure}[t]
\centering
\includegraphics[width=0.5 \linewidth] {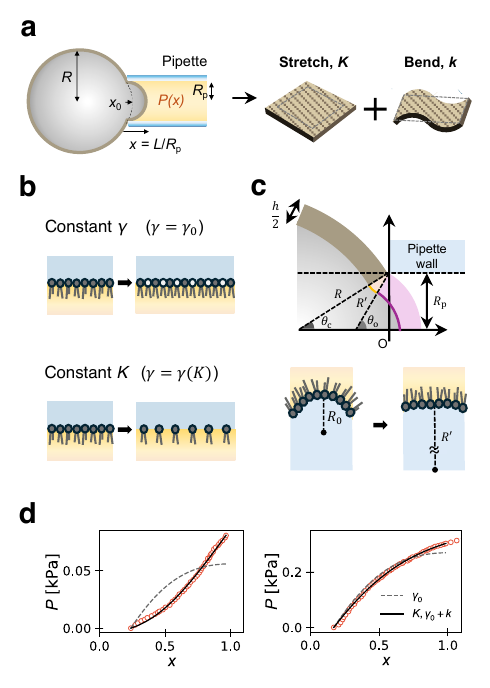}
\caption{
\textbf{Analytical model for the deformation of membrane-covered spherical droplets during micropipette aspiration.} 
(a) A droplet with radius $R$ is aspirated by a pipette with an inner radius $R_\mathrm{p}$ under a certain pressure difference $P$ between the external environment and the interior of the pipette. The resulting deformation in the aspirated region includes the surface stretching and bending due to the aspiration length $L$. Stretching and bending are characterized by an area-stretching modulus $K$ and a bending modulus $k$, respectively.  
(b) Two different molecular-level scenarios of the interface during surface stretching: (Top) When the lipid supply rate from the reservoir is sufficiently fast compared to deformation time, the interfacial tension $\gamma$ is maintained to be constant ($\gamma$=$\gamma_\mathrm{0}$), and (Bottom) when the supply rate is slow, the strain is constant (constant area-stretching modulus, $K$), but $\gamma$ varies with expansion ($\gamma$=$\gamma$($K$)). 
(c) Curvature change of the lipid membrane: (Top) The schismatic of the neutral surface of the lipid membrane 
with a thickness $h$ in the aspirated region inside the pipette. Curvature variations mainly arise in the aspirated spherical segment (purple) and the pipette edge (yellow). 
(Bottom) Considering the spontaneous curvature of lipids (curvature radius $R_0$), curvature changes from the initial $1/R_0$ to $1/R'$ in the aspirated spherical segment. 
(d) Two typical experimentally obtained $P$--$x$ curves (red; left: convex-downward, right: convex-upward) and their fits using our model based on Equation~\ref{eq:model} (black solid lines) and the conventional fits with a constant $\gamma_\mathrm{0}$ (black dashed lines).
}
\label{fig:1} 
\end{figure}

To analyze the interfacial mechanical properties of spherical droplets coated with a lipid monolayer, which serve as simple synthetic cells, we employed a classical pipette aspiration setup, as illustrated in Figure~\ref{fig:1}a. In this configuration, a pressure difference $P$ between the interior and exterior of the pipette with an inner radius $R_{\mathrm{p}}$, deforms the membrane, pulling it into the pipette by a length $L$. We focus on the regime of small deformations, where the aspiration length is normalized by the pipette radius,
\begin{equation}
x \equiv \frac{L}{R_p},
\end{equation}
remains below unity. In this regime ($x<1$), deformations outside the pipette can be
neglected (see Figure~S1 in the Supporting Information), and the mechanical response is governed primarily by the energies of the aspirated region.

To analyze the experimentally obtained nonlinear $P$--$x$ curves, we consider the change in elastic energy associated with aspiration, which consists of the area-stretching energy due to changes in the membrane surface area, $\Delta E_{\mathrm{stretch}}$, and the bending
energy arising from curvature variations, $\Delta E_{\mathrm{bend}}$:
\begin{equation}
\Delta E_{\mathrm{stretch}} + \Delta E_{\mathrm{bend}}
= \int_{x_0}^{x} P_{\mathrm{diff}}(x)\, \frac{\mathrm{d}V}{\mathrm{d}x}\, \mathrm{d}x .
\label{eq:energy-pressure}
\end{equation}
Here, $P_{\mathrm{diff}}(x)$ denotes the effective pressure difference, which will be
described in detail below. The parameter $x_0$ is the initial normalized aspiration
length originating from the intrinsic curvature of the droplet.

The volume $V$ and surface area $A$ of the aspirated region are given geometrically as
(see Section~S1 for details):
\begin{align}
V &= \frac{\pi R_p^3}{6}\left( 3x + x^3 \right), \\
A &= \pi R_p^2 \left( 1 + x^2 \right).
\end{align}

\subsection{Area-stretching energy}

When considering area stretching of a lipid monolayer membrane, two distinct molecular-level scenarios arise depending on the rate at which lipids are supplied from the reservoir, as illustrated in Figure~\ref{fig:1}b. 
If the characteristic timescale of aspiration-induced deformation is sufficiently long compared to the lipid supply timescale, the newly created interface is rapidly populated by lipid molecules and reaches a quasi-equilibrium state. Under this condition, the interfacial tension remains constant at its equilibrium value, $\gamma = \gamma_0$ (Figure~\ref{fig:1}b, upper). In contrast, when the deformation occurs on a shorter timescale, the number of lipid molecules that can adsorb to the expanding interface during deformation is limited. As a result, $\gamma$ cannot equilibrate instantaneously and instead evolves dynamically with time (Figure~\ref{fig:1}b, lower).

For synthetic cells such as droplets and liposomes, the supply of lipids typically occurs on timescales longer than $\sim 1~\mathrm{s}$ \cite{Miller1998}, which is comparable to the aspiration timescale used in this study. We therefore adopt the latter scenario with a
time-varying interfacial tension.

Assuming uniform strain in the aspirated region, the area-stretching modulus $K$ is treated as constant, and the interfacial tension $\gamma$ is expressed as a function of the areal strain
\begin{equation}
\varepsilon_A \equiv \frac{A - A_0}{A_0},
\end{equation}
as
\begin{equation}
\gamma(x) = \gamma_0 + K \varepsilon_A
          = (\gamma_0 - K) + K \frac{A}{A_0},
\end{equation}
where $\gamma_0$ and $A_0$ denote the interfacial tension and area prior to aspiration,
respectively.

The area-stretching energy is then calculated by integrating $\gamma$ over the aspirated
surface area:
\begin{equation}
E_{\mathrm{stretch}}
= \int \gamma\, \mathrm{d}A
= (\gamma_0 - K)A + \frac{K}{2}\frac{A^2}{A_0}.
\label{eq:E_stretch}
\end{equation}

\subsection{Bending energy}

According to the classical description of membrane bending rigidity \cite{Helfrich1973}, the bending energy associated with curvature deformations is given by
\begin{equation}
E_{\mathrm{bend}} = \int 2k \left( H - c_0 \right)^2 \mathrm{d}A ,
\end{equation}
where $k$ is the bending modulus, $H$ is the mean curvature, and $c_0$ is the spontaneous
curvature of the membrane.

In our lipid monolayer system, $c_0$ is primarily determined by the molecular geometry
of the lipids coating the droplet. During pipette aspiration, curvature variations
mainly arise in two regions: the aspirated spherical segment inside the pipette and the pipette edge (Figure~\ref{fig:1}c).

For the aspirated spherical region, the radius of curvature is
\begin{equation}
R' = R_p \frac{1 + x^2}{2x}.
\end{equation}
The corresponding bending energy is
\begin{align}
E_{\mathrm{sphere}}
&= 2k\, A \left( \frac{1}{R'} - \frac{1}{R_0} \right)^2 \nonumber \\
&\approx \frac{\pi k}{2} \left( \frac{R_p}{R_0} \right)^2 (1 + x^2)
= \mathcal{O}\!\left( \frac{R_p^2}{R_0^2} \right),
\label{eq:E_sphere}
\end{align}
where $R_0$ is the intrinsic curvature radius defined by $c_0 = 1/R_0$.

At the pipette edge, the membrane curvature is much larger than that of the spherical regions and is governed by the monolayer thickness $h \ll R_p$. 
Approximating this junction as an axisymmetric surface of revolution with a characteristic curvature of order $(h/2)^{-1}$ (Figure~\ref{fig:1}c), and using the geometric condition $h, R_0 \ll R'$,
the bending energy at the edge is estimated as (see Section~S2)
\begin{align}
E_{\mathrm{edge}}
&\approx 2k A_{\mathrm{edge}}
\left[
\frac{1}{2}\left( \frac{1}{h/2} - \frac{1}{R_p} \right) - \frac{1}{R_0}
\right]^2 \nonumber \\
&\approx \frac{k}{2} A_{\mathrm{edge}}
\left( \frac{1}{h/2} - \frac{2}{R_0} \right)^2
= \mathcal{O}\!\left( \frac{R_p}{h} \right).
\end{align}
Because $h$ and $R_0$ are on the nanometer scale, whereas $R_p$ is on the micrometer
scale, we obtain $E_{\mathrm{edge}} \ll E_{\mathrm{sphere}}$. The contribution from
the pipette edge can therefore be neglected.

\subsection{Expression of experimentally obtained force curve}

From Equations~\ref{eq:energy-pressure}, \ref{eq:E_stretch}, and \ref{eq:E_sphere}, the effective pressure difference between the interior of the pipette and that of the droplet is given by
\begin{equation}
P_{\mathrm{diff}}(x)
= \frac{4}{R_p}
\left[
\frac{K}{1 + x_0^2} x
+ (\gamma_0 - K)\frac{x}{1 + x^2}
\right]
+ \frac{2k}{R_p R_0^2} \frac{x}{1 + x^2}.
\label{eq:P_diff}
\end{equation}

To compare with the experimentally obtained $P$--$x$ curve, the effective pressure difference $P_{\mathrm{diff}}(x)$ was adjusted by subtracting the contributions from interfacial tension, referred to as Young–Laplace pressure, and from the restoring pressure resulting from droplet bending outside the pipette \cite{Fournier2008,Rangamani2014} (see Section~S3):
\begin{equation}
P(x) = P_{\mathrm{diff}}(x)
- \left[ \frac{2\gamma(x)}{R} + \frac{k}{R}\left( \frac{1}{R_0} \right)^2 \right].
\label{eq:modified_YL}
\end{equation}

Combining Equations~\ref{eq:P_diff} and \ref{eq:modified_YL}, we finally obtain
\begin{equation}
P(x)
= \frac{4}{R_p}
\left(
\gamma_0 + \frac{k}{2R_0^2}
+ K \frac{x^2 - x_0^2}{1 + x_0^2}
\right)
\left(
\frac{x}{1 + x^2} - \frac{R_p}{2R}
\right),
\label{eq:model}
\end{equation}
which contains two fitting parameters,
$\gamma_0 + k/(2R_0^2) \equiv \gamma_0 + \tilde{k}$ and $K$.
Depending on these parameters, the model reproduces both convex-downward and
convex-upward $P$--$x$ curves (Figure~\ref{fig:1}d and Figure~S2).

\section{Results and Discussion}
\subsection{Experimental Validation of the Analytical Model}

Before investigating the effects of lipids and DNA nanostructures on the interfacial mechanics of synthetic cells, we first validated the proposed analytical model (Equation~\ref{eq:model}) using droplets coated with a lipid monolayer. These droplets provide a well-defined and controllable model system, enabling direct comparison between experimental measurements and theoretical predictions. Figure~\ref{fig:1}d displays two representative pressure--displacement ($P$--$x$) curves obtained from pipette aspiration, exhibiting either downward-convex (upper) or upward-convex (lower) behavior. The analytical model (solid lines) quantitatively reproduces the experimental data over the entire displacement range for both response types ($R^2 = 0.99 \pm 0.03, \text{mean} \pm \text{s.d.}$). Notably, the model captures both convex and concave nonlinearities within a single unified framework, a capability that conventional approaches cannot achieve.
In contrast, the conventional Young--Laplace fitting assumes a constant interfacial tension $\gamma_0$ (dashed lines) and shows systematic deviations. This is especially apparent for downward-convex responses and at larger displacements ($R^2 = 0.94 \pm 0.07, \text{mean} \pm \text{s.d.}$). Neglecting the elastic contribution in the modified Young--Laplace pressure (Equation~\ref{eq:modified_YL}) further worsens the fitting quality. This result underscores the necessity of incorporating elastic effects. Together, these results confirm the validity of the proposed analytical framework and establish a reliable basis for subsequent analysis of how lipid composition and DNA nanostructure modulate the mechanical properties of droplets as synthetic cell models.

\subsection{Lipid-Dependent Interfacial Mechanics of Synthetic Cell Droplets}

\begin{figure}[t]
\centering
\includegraphics[width=1 \linewidth] {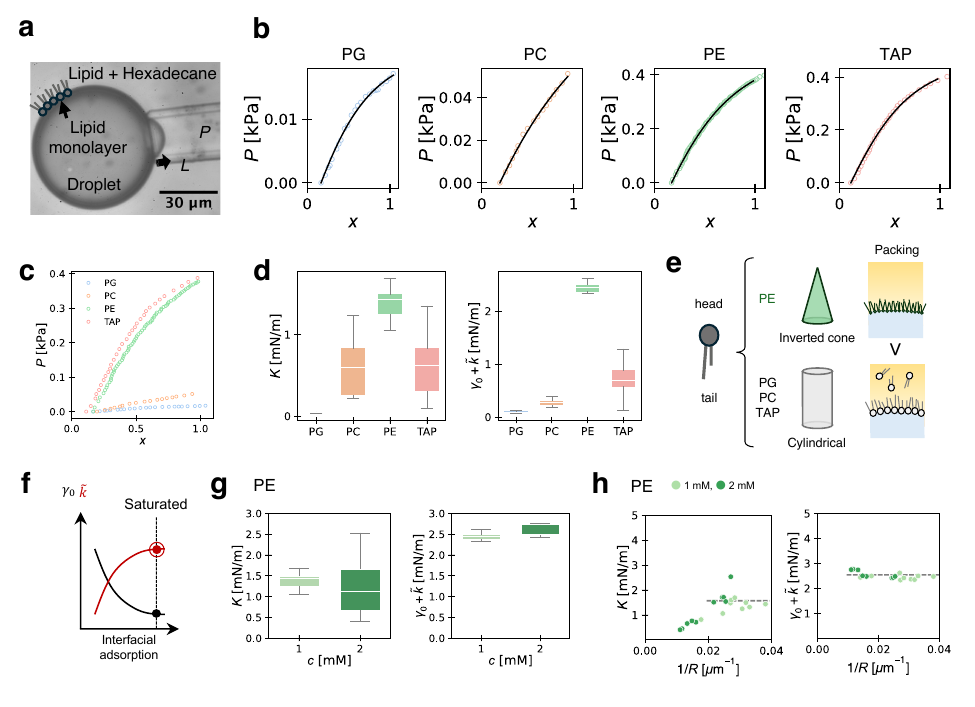}
\caption{
\textbf{Micropipette aspiration of droplets with different types of lipids.}
(a) A microscope image showing a droplet aspirated by a pipette with $R_{\mathrm{p}}=12$ $\mu$m at an aspiration pressure, $P$. 
(b, c) Typical $P$--$x$ curves for droplets with each lipid (circles) and their fits using our model based on Equation~\ref{eq:model} (black solid lines). 
(d) $K$ (left) and $\gamma_0+\tilde{k}$ (right) for each lipid droplets. PE shows the largest values both in $K$ and $\gamma_0+\tilde{k}$. ($n$ = 6, 7, 9 and 25 for PG, PC, PE and TAP, respectively).
(e) Schematics illustrating lipid molecular geometry and their adsorption at the interface. PE has a relatively small headgroup cross-sectional area compared with its hydrophobic tails, i.e. inverted-cone geometry. PC, PG, and TAP possess comparable head and tail cross-sectional areas, i.e. cylindrical lipids. Upon adsorption onto a surface with a large radius of curvature, PE molecules are easily packed on the membrane with small headgroups, resulting in an increased surface number density compared with the other lipids.
(f) Schematic illustration of the opposite dependence of $\gamma_0$ and $\widetilde{k}$ on interfacial adsorption. While increasing interfacial adsorption generally reduces $\gamma_0$, it simultaneously enhances $\widetilde{k}$. When the spontaneous curvature $c_0$ of the adsorbed molecules is larger, $\widetilde{k}$ is shifted to higher values.
(g) $K$ (left) and $\gamma_0+\tilde{k}$ (right) for PE droplets at two different lipid concentrations. The absence of any significant change in either parameter indicates that the interfacial adsorption of PE is already saturated at 1 mM.
(h) The curvature dependency of $K$ (left) and $\gamma_0+\tilde{k}$ (right) for PE droplets. Light gray dashed lines indicate the average values. The dashed line for $K$ is calculated using data with $1/R \gtrsim 0.02~\mu\mathrm{m}^{-1}$.
}
\label{fig:2} 
\end{figure}

To compare the interfacial mechanics of lipid-monolayer-coated droplets, we examined four lipids with distinct headgroup charges: PG (negatively charged), PC and PE (electrically neutral), and TAP (positively charged) (see Experimental section for details). Aspiration of the droplets yielded the $P$--$x$ curves for each lipid, as presented in Figure~\ref{fig:2}a and b. These nonlinear curves were obtained at the regime of relatively small deformation that precedes the onset of viscous flow in the large-deformation region. Regardless of headgroup charge, all lipids exhibit upward convex $P$--$x$ curves that are accurately described by the model using Equation~\ref{eq:model}.

Under identical aspiration lengths $x$, droplets coated with PE (green) and TAP (pink) required substantially higher aspiration pressures than those coated with PC (orange) and PG (blue) (Figure~\ref{fig:2}c), revealing pronounced lipid-\\dependent differences in interfacial mechanical response. To quantify these differences, we fitted each $P$--$x$ curve using Equation~\ref{eq:model} and extracted the area-stretching modulus $K$ and the bending-related term $\gamma_0+\tilde{k}$. Both parameters were largest for PE (electrically neutral), followed by TAP (positively charged), PC (electrically neutral), and PG (negatively charged) (Figure~\ref{fig:2}d).
The magnitude of the extracted area-stretching modulus $K$ is consistent with values reported for related lipid-coated interfaces. For example, oscillatory deformation measurements of oil–water interfaces stabilized by a lipid monolayer have reported $K$ values on the order of $1~\mathrm{mN\,m^{-1}}$ \cite{Jin2021}, comparable to those obtained here. 
This agreement support the physical validity of the extracted parameters and demonstrates that the present model yields quantitatively reasonable mechanical moduli.

The systematic ordering of the two extracted mechanical parameters, with PE exhibiting markedly higher values than the other lipids, cannot be attributed to headgroup charge but instead can reflect intrinsic differences in lipid molecular geometry. Such geometry is commonly characterized by the spontaneous curvature $c_0$, which arises from a mismatch between the cross-sectional areas of the lipid headgroup and hydrophobic tail. As illustrated in Figure~\ref{fig:2}e, cylindrical lipids such as PC, PG, and TAP possess near-zero spontaneous curvature and therefore favor planar monolayer configurations. In contrast, PE has an inverted-cone molecular shape with a negative spontaneous curvature ($c_0<0$)~\cite{Dymond2021}, originating from its relatively small headgroup compared to the hydrophobic tail, and thus intrinsically prefers membranes curved toward the hydrophilic side.

We first address why PE exhibits a substantially larger $K$ than the other lipids, focusing on molecular geometry, spontaneous curvature, and the resulting in-plane packing of the lipid monolayer. The modulus $K$ quantifies the free-energy penalty associated with lateral area expansion of the membrane, which requires separation of neighboring lipid molecules within the interfacial plane. Lipid planer membranes with stronger in-plane cohesion and higher packing density, therefore, exhibit larger values of $K$ \cite{Den2007}.
Such packing characteristics are intrinsically linked to lipid molecular geometry, as encoded by the spontaneous curvature $c_0$. In contrast to cylindrical lipids with near-zero $c_0$, inverted cone-shaped lipids with non-zero spontaneous curvature, such as PE, tend to pack more densely within the interfacial plane when constrained to form a curved monolayer (Figure~\ref{fig:2}e). This enhanced packing on the curved membrane leads to a higher energetic cost upon lateral expansion, resulting in an increased $K$. Consistent with this interpretation, cone-shaped lipids with non-zero spontaneous curvature, including PE, have been reported to exhibit enhanced in-plane cohesion and ordering compared with cylindrical lipids, as evidenced by increased rigidity in Langmuir monolayer compressibility measurements and interfacial rheology studies \cite{Manna2021,Luna2011}.
These results demonstrate that, even in three-dimensionally closed membranes such as synthetic cells, in-plane lipid packing, which depends on lipid geometry, plays a dominant role in governing resistance to lateral deformation.

We next examine why PE also exhibits a substantially larger value of $\gamma_0+\tilde{k}$ compared with the other lipids. The first contribution, the interfacial tension $\gamma_0$, represents the free energy of the oil--water interface covered by a lipid monolayer. In general, adsorption of amphiphilic molecules reduces $\gamma_0$ as the interfacial adsorption density increases. In contrast, the bending-related contribution $\tilde{k}$ has been reported to increase with increasing adsorption density, reflecting enhanced resistance to curvature deformation \cite{Seto2006, Rekvig2004}. As a result, the composite parameter $\gamma_0+\tilde{k}$ is expected to exhibit a weak or negligible dependence on the lipid adsorption density due to these opposing trends and $\tilde{k}$ is maximized under lipid saturation~(Figure~\ref{fig:2}f).

To test whether the markedly larger $\gamma_0+\tilde{k}$ observed for PE arises from incomplete interfacial saturation, we increased the lipid concentration in the oil phase from 1~mM (above the critical micelle concentration) to 2~mM. No statistically significant changes were detected in either fitted parameters (Figure~\ref{fig:2}g), indicating that the oil--water interface is already saturated at 1~mM under the present conditions. Independent interfacial tension measurements using the Wilhelmy plate method have reported that PE exhibits a higher equilibrium tension $\gamma_0$ than cylindrical lipids \cite{Yanagisawa2013}, consistent with a dominant contribution from its intrinsic molecular geometry.

We further analyzed the droplet-size dependence to disentangle the relative contributions of interfacial tension and bending elasticity (Figure~\ref{fig:2}h). The composite parameter $\gamma_0+\tilde{k}$ remains nearly independent of $1/R$ over the entire size range investigated. This size invariance suggests that the changes in $\gamma_0$ and $\tilde{k}$ caused by the shift in effective adsorption packing compensate each other as the droplet curvature changes. In contrast, the area-stretching modulus $K$ of PE exhibits a pronounced size dependence: while $K$ remains constant for $1/R \gtrsim 0.02~\mu\mathrm{m}^{-1}$ (corresponding to $R \lesssim 50~\mu\mathrm{m}$), it decreases systematically for larger droplets with lower curvature (no comparable size dependence for the other lipids; Figure~S3) .

This behavior can be rationalized by considering the intrinsic negative spontaneous curvature of PE. For $\mu\mathrm{m}$-sized droplets, the membrane surface appears nearly planar on the molecular length scale. As a result, PE molecules, which favor curved configurations ($c_0<0$), experience packing frustration on flatter, larger droplets, leading to a reduced packing density and a smaller apparent $K$. Although such reduced packing would tend to increase $\gamma_0$ due to a lower effective adsorption packing density, it concurrently reduces the bending contribution $\tilde{k}$, yielding an approximately constant value of $\gamma_0+\tilde{k}$. These findings indicate that the enhanced $\gamma_0+\tilde{k}$ observed for PE is primarily due to the bending contribution of PE membrane under highly packed conditions resulting from its inverted-cone shape. In addition, the inverted-cone shape of PE exhibits stable and well-defined mechanical behavior only above a threshold curvature, which under the present conditions corresponds to $1/R \gtrsim 0.02~\mu\mathrm{m}^{-1}$, since inefficient packing limits membrane integrity at the low-curvature interfaces of large droplets.

\subsection{DNA-Nanostructure-Induced Modulation of Interfacial Mechanics}

\begin{figure}[t]
\centering
\includegraphics[width=1 \linewidth] {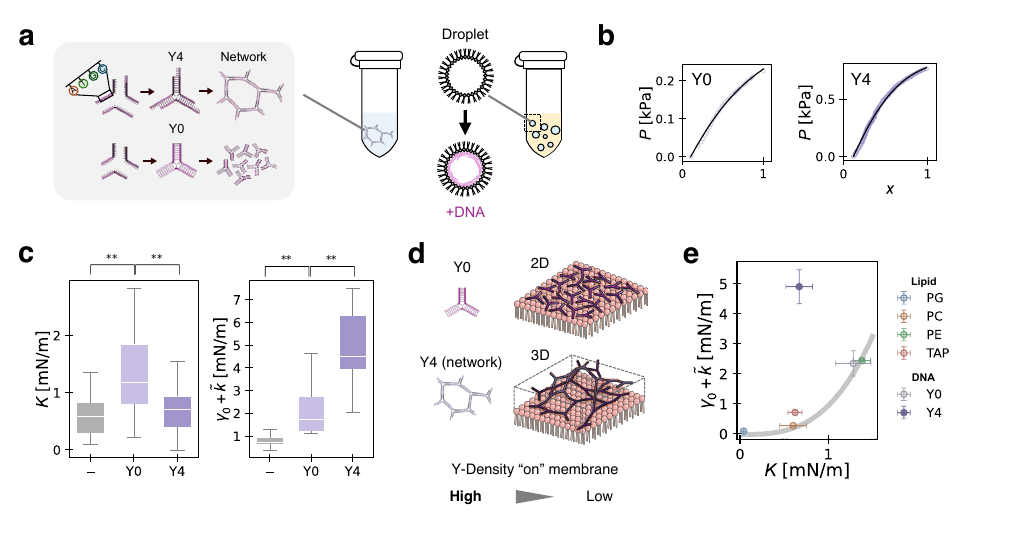}
\caption{
(a) Schematics of hybridization of DNA with and without sticky ends by decreasing the temperature. The three different DNA oligomers form a Y-shaped dsDNA motif. In the presence of sticky ends, the Y-motif DNAs further hybridize to create a network structure, whereas motifs without sticky ends remain isolated. These nanostructures electrostatically localize beneath the membrane when encapsulated within lipid-coated droplets.
(b) Typical $P$--$x$ curves for droplets with DNA nanostructures (circles) and their fits using our model based on Equation~\ref{eq:model} (black solid lines). 
(c) $K$ (left) and $\gamma_0+\tilde{k}$ (right) for each DNA nanostructures (Y0 and Y4). For reference, the DNA-free values (--) obtained for 1~mM TAP in the previous section are also shown ($n$ = 25, 16 and 10 for --, Y0, and Y4, respectively).
(d) Schematics of Y0 and Y4, which are confined in quasi-two-dimensional and three-dimensional configurations, respectively. Despite having the same total number of DNA motifs, the effective surface density of DNA motifs directly adsorbed at the interface is higher for Y0 than for Y4.
(e) The average values of $K$ and $\gamma_0+\tilde{k}$ for Y0 and Y4 with those of different lipids (obtained from Figure~\ref{fig:2}d). The error bars represent the standard error. A light gray line indicates the scaling relation, $\propto K^3$, expected from thin-film elasticity theory.
}
\label{fig:3} 
\end{figure}

To quantitatively assess how the adsorption of DNA nanostructures modulates the interfacial mechanics of droplets, we encapsulated self-assembled DNA motifs within droplets coated with a cationic TAP monolayer and performed micropipette aspiration experiments. Three partially complementary DNA strands were annealed to form Y-shaped motifs. By introducing terminal sticky ends, we generated cross-linked networks (Y4) and compared them with isolated motifs without the sticky ends (Y0, Figure~\ref{fig:3}a). Owing to their negative charge, the DNA motifs are electrostatically attracted to the positively charged TAP monolayer and localize beneath the droplet interface \cite{Kurokawa2017}.

At identical aspiration lengths $x$, droplets containing Y4 required higher aspiration pressures than those containing Y0 (Figure~\ref{fig:3}b), indicating that DNA's network formation enhances interfacial resistance to deformation. To quantify these effects, the pressure--displacement curves were fitted using Equation~\ref{eq:model}, yielding the area-stretching modulus $K$ and the bending-related term $\gamma_0+\tilde{k}$. The modulus $K$ was highest for Y0, whereas Y4 and the DNA-free control exhibited comparable values (Figure~\ref{fig:3}c, left). In contrast, $\gamma_0+\tilde{k}$ increased monotonically from the DNA-free condition (1~mM TAP) to Y0 and was largest for Y4 (Figure~\ref{fig:3}c, right). It should be noted that the aqueous phase here contained Tris--HCl buffer rather than the NaCl solution used droplets without DNA (shown in Figure~\ref{fig:2}).

We first consider the behavior of $K$. Comparison between Y0-containing and DNA-free droplets shows that adsorption of DNA motifs significantly enhances the in-plane stiffness of the monolayer. This result is notable, as reinforcement of membrane mechanics in both biological and synthetic cells typically relies on polymeric or cytoskeleton networks \cite{Daly2024, Ganar2021}. Here, isolated DNA motifs increase lateral resistance solely through interfacial adsorption, without forming an extended network. The lower $K$ observed for Y4 compared with Y0 indicates that sticky-end-mediated crosslinking reduces the effective surface density of DNA motifs directly adsorbed at the interface. Although the Y4 motifs are electrostatically localized near the membrane, their preassembled three-dimensional network structure prevents efficient flattening into a quasi-two-dimensional layer (Figure~\ref{fig:3}d). With an interaction energy on the order of $\sim10\,k_\mathrm{B}T$ per junction and a mesh size of $\sim20$~nm, flattening the network would require substantial deformation or rupture of multiple cross-links, which is energetically unfavorable. As a result, only a fraction of motifs can directly contribute to area stretching, leading to a reduced apparent $K$. This mechanism parallels the lipid-dependent behavior discussed above, in which reducing the effective packing density decreases the magnitude of $K$.

We next consider the bending-related term $\gamma_0+\tilde{k}$. Comparison between Y0 and DNA-free droplets shows that adsorption of DNA motifs substantially increases $\gamma_0+\tilde{k}$, i. e., enhancing resistance to out-of-plane deformation. Because the persistence length of double-stranded DNA ($\sim50$~nm) exceeds the characteristic dimensions of the DNA motifs, the adsorbed DNA motifs behave as rigid inclusions that enhance the bending rigidity of the interface. Notably, $\gamma_0 + \widetilde{k}$ further increased for Y4 compared with Y0, despite a reduction in the packing density inferred from the lower $K$ of Y4 than Y0. This result highlights that $\gamma_0 + \widetilde{k}$ for Y4 is predominantly governed by the bending $\widetilde{k}$ rather than the tension contribution. In other words, while $K$ primarily reflects the two-dimensional packing density of DNA motifs directly adsorbed at the interface, $\widetilde{k}$ is governed by the mechanical rigidity and finite thickness of the three-dimensional DNA network located beneath the lipid membrane.

To derive the mechanical properties of the three-dimensional structure of the Y4 DNA on the lipid membrane in more detail, we plotted the correlation between the two extracted parameters, and compared them with those of various two-dimensional systems: lipid-only membrane (the same data shown in Figure 2d) and a lipid membrane plus Y0 DNA (Figure ~\ref{fig:3}e). According to thin film theory, the bending rigidity $k$ is proportional to the cube of the membrane thickness, while the areal stretching coefficient $K$ is proportional to the thickness \cite{Landau2012}. Therefore, excluding thickness, we obtain a scaling relationship of the form $k \propto K^3$. For all systems except Y4 DNA, the measured mean values align with a single scaling curve proportional to $K^3$ (indicated by a gray line in Figure~\ref{fig:3}e). This observation indicates that the mechanically effective membrane thickness remains approximately constant across these lipid systems, regardless of the presence of Y0 DNA. Additionally, the results indicate that the elastic contribution $\widetilde{k}$ predominates in the composite parameter $\gamma_0 + \widetilde{k}$, as noted above. Thus, for droplets coated with spontaneously adsorbed elements, the interfacial mechanics are governed primarily by the packing density of elements directly adsorbed at the membrane. In contrast, Y4 forms a three-dimensional DNA network, resulting in an effective membrane thickness that is significantly greater than that of lipid-only membranes or those decorated with Y0. Consistent with this expectation, the Y4 data points deviate markedly from the common scaling curve.

These results demonstrate that even for identical DNA nanostructures, the mechanical contribution to the interface depends fundamentally on whether they exist as isolated motifs (Y0) or as a crosslinked three-dimensional network (Y4). The former primarily enhances resistance to area stretching, whereas the latter, by acting as a finite-thickness structure, dominantly increases resistance to bending deformations. Moreover, our findings indicate that introducing three-dimensional network architectures into synthetic cells is an effective strategy for selectively enhancing bending resistance in the direction normal to the interface, as reflected by an increased $\widetilde{k}$.

Finally, we estimate the bending rigidity $k$ of droplets bearing a three-dimensional Y4 DNA network beneath the lipid membrane. The spontaneous curvature of PE has been reported to lie in the range $c_0 \simeq -0.05$ to $-0.03~\mathrm{nm}^{-1}$ \cite{Dymond2021}; here, we adopt a representative value of $c_0 = -0.04~\mathrm{nm}^{-1}$. From the experimental fitting, the bending-related term $\tilde{k}$ is on the order of $\sim 1~\mathrm{mN\,m^{-1}}$. Using the relation $\tilde{k} = k/(2R_0^2) = kc_0^2/2$, we estimate the bending rigidity of droplets with the Y4 DNA network to be $k \sim 10^2\,k_{\mathrm{B}}T$. This value is approximately five times larger than that obtained for droplets without DNA or with isolated Y0 DNA motifs (Figure ~\ref{fig:3}e), indicating a substantial enhancement in bending rigidity induced by the three-dimensional DNA network. Notably, the estimated magnitude is comparable to reported bending rigidities of red blood cell membranes, which span from a few to $\sim 10^2\,k_{\mathrm{B}}T$ depending on measurement conditions \cite{Himbert2022}. This quantitative agreement underscores that adsorption of a finite-thickness DNA network beneath a lipid membrane can endow synthetic cells with bending mechanics approaching those of biological membranes.

\section{Conclusion}
In this study, we established a quantitative, unified framework for describing the nonlinear interfacial mechanics of lipid-membrane-coated droplets, serving as synthetic cells, by combining systematic micropipette aspiration experiments with a physically based analytical model. By explicitly incorporating both area stretching and bending rigidity into the description of aspiration deformation, the proposed model quantitatively reproduces the full pressure–displacement ($P$–$x$) response in the small but nonlinear regime preceding viscous flow. This combined experimental–theoretical approach reduces interfacial mechanics to two physically interpretable parameters: the area-stretching modulus $K$, which captures in-plane resistance, and a composite interfacial-tension and bending-related term, $\gamma_0+\tilde{k}$, governing out-of-plane deformation.

Using these experimentally accessible parameters as a common mechanical metric, we demonstrate that interfacial mechanics across chemically distinct systems can be understood within a shared physical framework. For lipid monolayers, intrinsic molecular curvature and packing constraints jointly regulate mechanical response. Inverted-cone-shaped PE, characterized by a large negative spontaneous curvature, exhibits substantially enhanced values of both $K$ and $\gamma_0+\tilde{k}$ compared with cylindrical lipids such as PC, PG, and TAP. Concentration- and size-dependent measurements, made interpretable by the analytical model, further reveal that interfacial packing selectively modulates $K$, whereas $\gamma_0+\tilde{k}$ remains largely insensitive to droplet size, consistent with its bending contribution.

Extending this experimentally validated framework to DNA-functionalized interfaces reveals an additional, orthogonal mechanism for mechanical control that arises from dimensionality. Isolated DNA motifs (Y0) behave as quasi-two-dimensional adsorbates that primarily enhance in-plane stiffness by increasing $K$ through direct interfacial coupling. In contrast, cross-linked DNA nanostructures (Y4) form a finite-thickness three-dimensional network beneath the lipid monolayer, which is directly captured by the model as a pronounced increase in $\gamma_0+\tilde{k}$, reflecting enhanced resistance to bending even at reduced effective interfacial adsorption density.

Together, these results demonstrate that only by integrating quantitative experiments with an analytical description of nonlinear deformation can one directly link molecular-scale organization to emergent interfacial mechanics. This work establishes a general design principle for soft interfaces, in which in-plane and out-of-plane mechanical properties can be independently and predictably programmed by controlling molecular geometry, effective surface density, and the dimensionality of adsorbed components. By providing a unified physical framework that connects molecular design to nonlinear mechanical function, our study lays the foundation for the rational engineering of mechanically programmable synthetic cells, membrane-based soft materials, and biomimetic systems whose functionality is encoded in interfacial mechanics.

\section{Materials and Methods}
\subsection{Materials}
1,2-dioleoyl-sn-glycero-3-phosphoglycerol (PG), 1,2-dioleoyl-sn-glycero-3-phosphocholine (PC), 1,2-dioleoyl-sn-\\glycero-3-phosphoethanolamine (PE), and 1,2-dioleoyl-3-trimethylammonium-propane (TAP; chloride salt) were purchased from Avanti Polar Lipids. Lipids were dissolved in chloroform (Nacalai Tesque), and lipid films obtained after solvent evaporation were redissolved in hexadecane (Nacalai Tesque) as the oil phase. Sodium chloride (NaCl; FUJIFILM Wako Pure Chemical Co.) and Tris–HCl buffer (pH 8.0; Nacalai Tesque) were dissolved in ultrapure water (Invitrogen) to prepare the aqueous phase. For experiments involving DNA motifs, lyophilized DNA oligonucleotides (salt-free grade; Eurofins Genomics Japan) were dissolved in ultrapure water and added to the aqueous phase as appropriate. DNA stock solutions were prepared at 2 mM and stored at -20~$^\circ$C until use. All materials were used as received without further purification.

\subsection{Droplet preparation}
Water-in-oil droplets encapsulated by lipid monolayers were prepared as follows. Lipids were first dissolved in chloroform at a final concentration of 10 mM. An aliquot (50 $\mu$L) of the lipid solution was transferred to a Durham tube, and the solvent was evaporated under a gentle stream of nitrogen gas to form a dry lipid film at the bottom of the tube. Hexadecane (500 $\mu$L) was then added, and the mixture was sonicated at approximately 60~$^\circ$C for 90 min to obtain a lipid-in-oil solution with a final lipid concentration of 1 mM. The solution was subsequently cooled slowly from 60~$^\circ$C to room temperature (approximately 25~$^\circ$C) with intermittent vortex mixing to ensure homogeneous lipid dissolution.

The aqueous phase consisted of a 370 mM NaCl solution, unless otherwise noted. For experiments involving DNA nanostructures, the composition of the aqueous phase was modified as described in the following section. To generate droplets, 2 $\mu$L of the aqueous phase was added to 40 $\mu$L of the lipid-in-hexadecane solution after sonication, and the mixture was gently tapped to induce droplet formation without bulk emulsification. The resulting droplets were transferred onto a silicone-coated glass-bottom dish (Matsunami) to prevent adhesion to the glass surface and were used for subsequent observations and measurements.

\subsection{DNA nanostructure preparation}
DNA oligomers were designed based on a previously reported architecture to assemble into Y-shaped DNA nanostructures \cite{Kurokawa2017}. At high temperatures, the DNA oligomers remain dissociated due to entropic stabilization, whereas upon cooling they hybridize via Watson–Crick base pairing to form Y-shaped motifs. Two types of DNA nanostructures were prepared: motifs bearing sticky ends at the three termini (G1–G3) and motifs without sticky ends (S1–S3) (Figure~S4). Upon further cooling, the sticky-end–bearing nanostructures undergo additional hybridization between complementary sticky ends, resulting in the formation of extended network structures, whereas nanostructures without sticky ends do not form networks.

DNA nanostructures with sticky ends (G1–G3) exhibit two characteristic melting transitions. Specifically, Y-shaped motifs are formed below the first melting temperature, $T_{\mathrm{m1}}$, followed by inter-motif hybridization via sticky ends at the second melting temperature, $T_{\mathrm{m2}}$. In contrast, nanostructures without sticky ends (S1–S3) display a single melting transition at $T_{\mathrm{m1}}$ and remain as isolated motifs. The melting temperatures were $T_{\mathrm{m1}} = 69~^\circ$C and $T_{\mathrm{m2}} = 49~^\circ$C.

For self-assembly, the constituent DNA oligomers were mixed at equimolar ratios and diluted in buffer (20 mM Tris–HCl, pH 8.0, 350 mM NaCl) to a final DNA concentration of approximately 5 $\mu$M. The solution was heated to 80~$^\circ$C for 10 min and then slowly cooled to 10~$^\circ$C at a rate of 0.01~$^\circ$C,s$^{-1}$ using a thermal cycler (T-Gradient; Biometra), allowing the DNA nanostructures to self-assemble in a controlled manner. To prepare artificial cells with DNA nanostructures localized beneath the membrane, 2 µL of the aqueous solution containing DNA nanostructures was added to 40 $\mu$L of the TAP-in-hexadecane solution after sonication. Owing to electrostatic interactions between the negatively charged DNA nanostructures (hereafter referred to as DNA motifs) and the positively charged TAP monolayer, the DNA motifs were selectively localized beneath the lipid monolayer at the droplet interface.

\subsection{Micropipette aspiration}
The mechanical properties of artificial cells were quantified using the classical micropipette aspiration technique. In this method, a glass micropipette with an inner radius much smaller than that of the target droplet is brought into contact with the droplet surface, and the lipid monolayer is aspirated into the pipette by applying a pressure difference $P$ between the inside and outside of the pipette. The applied aspiration pressure induces a deformation of the interface, resulting in an aspiration length $L$ inside the pipette. The mechanical properties of the interface were estimated from the relationship between the aspiration pressure $P$ and the aspiration length $L$.

In this study, only droplets with radii $R \ge 20~\mu$m were analyzed, and glass micropipettes with an inner radius of $R_p \approx 15~\mu$m were used, corresponding to a geometric ratio of $R/R_p > 1.2$. Aspiration pressure and displacement were measured following a previously reported protocol \cite{Sakai2020}. Experiments were performed on an inverted optical microscope (Axiovert 40CFL; Carl Zeiss) equipped with a micromanipulator system (MMO-202ND and MN-4; Narishige), a microinjector (IM-11-2; Narishige), and a differential pressure transducer (DP15; Validyne). The droplet radius R and aspiration length L were determined from optical microscopy images.

\subsection{Statistical Analysis}
$P$–$x$ curves obtained from micropipette aspiration experiments were analyzed using nonlinear least-squares fitting based on the analytical model described in Equation 14. For each individual droplet, the aspiration region was identified automatically from the displacement time series, and fitting was restricted to the regime $x=L/R_{\mathrm{p}}\leq 1$, prior to the onset of relaxation or viscous flow. The fitting parameters, the area-stretching modulus $K$ and the combined interfacial tension–bending parameter $\gamma_0+\tilde{k}$, were obtained by minimizing the residual sum of squares using the trust-region reflective algorithm implemented in Python (SciPy, curve-fit). To avoid convergence to local minima, the fitting procedure was repeated with 10 randomly generated initial parameter guesses within physically meaningful bounds, and the solution yielding the highest coefficient of determination ($R^2$) was selected. Fits with $R^2<0.98$ were excluded from further analysis.

For statistical comparison of fitted parameters across different lipid compositions or DNA nanostructure conditions, droplets were treated as independent samples. Prior to group-level analysis, outliers were removed separately for each condition using the interquartile range (IQR) criterion (1.5×IQR). Data are presented as box plots with individual data points overlaid, and sample sizes ($n$) are indicated in the figure legends. Pairwise comparisons between groups were performed using Welch’s two-sided $t$-test, which does not assume equal variances. Statistical significance is indicated as $p<0.05$ and $p<0.01$, as specified in the figure captions. All data processing, fitting, and statistical analyses were conducted using custom Python scripts.

\section*{Data availability}
The data that support the findings of this study are available from the corresponding author upon reasonable request.

\section*{Conflicts of interest}
The authors declare no conflicts of interest.

\section*{Acknowledgements}
This research was partially funded by the Japan Society for the Promotion of Science (JSPS) KAKENHI (grant numbers  22H01188, 24H02287 (M.Y.), the Japan Science and Technology Agency (JST) (grant numbers FOREST, JPMJFR213Y; CREST (JPMJCR22E1) (M. Y.)), and the World-Leading Innovative Graduate Study Program for Advanced Basic Science Course (WINGS-ABC) at the University of Tokyo (K. M.).

\section*{Supporting information}
Supporting Information is available from the Wiley Online Library.

\bibliography{references}

\end{document}